%% file: sigir2021.tex
\documentclass[sigconf]{acmart}
\AtBeginDocument{%
  \providecommand\BibTeX{{%
    \normalfont B\kern-0.5em{\scshape i\kern-0.25em b}\kern-0.8em\TeX}}}

\setcopyright{acmcopyright}
\copyrightyear{2021}
\acmYear{2021}
\acmDOI{10.1145/1122445.1122456}

\acmConference[SIGIR '21]{Proceedings of the
44th International ACM SIGIR Conference
on Research and Development in Information Retrieval}{July 11--15, 2021}{Virtual Event, Canada}

\acmBooktitle{Proceedings of the 44th International ACM SIGIR Conference
on Research and Development in Information Retrieval, July 11--15, 2021, Virtual Event, Canada}
\acmPrice{15.00}
\acmISBN{978-1-4503-XXXX-X/18/06}

\usepackage{nicefrac}
\usepackage{multirow}
\usepackage{xspace}
\usepackage{booktabs}
\usepackage{makecell}
\usepackage{moresize}
\usepackage{tabularx}
\usepackage{adjustbox}
\usepackage[normalem]{ulem}
\usepackage{siunitx}
\usepackage{tikz}
\usepackage{amsthm}
\usepackage{graphicx}

\usepackage[show]{chato-notes}

\newcommand\boldparagraph[1]{\vspace{0.35em}\noindent\textbf{#1}}

\newcommand{\bmtf}{\textsf{BM25}\xspace}
\newcommand{\doctfquery}{\textsf{DocT5Query}}
\newcommand{\deepct}{\textsf{DeepCT}\xspace}
\newcommand{\hdct}{\textsf{HDCT}\xspace}
\newcommand{\colbert}{\textsf{ColBERT}\xspace}
\newcommand{\epic}{\textsf{EPIC}}
\newcommand{\prettr}{\textsf{PreTTR}}

\newcommand{\msmarcodev}{\textsf{MSMARCO Dev Queries}\xspace}
\newcommand{\trecdl}{\textsf{TREC 2019}\xspace}
\newcommand{\trecdltw}{\textsf{TREC 2020}\xspace}

\newcommand{\multibert}{\textsf{DeepImpact}}
\begin{document}
\fancyhead{}

\title{Learning Passage Impacts for Inverted Indexes}

\settopmatter{authorsperrow=4}
\author{Antonio Mallia}
\email{antonio.mallia@nyu.edu}
\affiliation{
  \institution{New York University}
  \country{}
}

\author{Omar Khattab}
\email{okhattab@stanford.edu}
\affiliation{
  \institution{Stanford University}
  \country{}
}
\author{Torsten Suel}
\email{torsten.suel@nyu.edu}
\affiliation{
  \institution{New York University}
  \country{}
}
\author{Nicola Tonellotto}
\email{nicola.tonellotto@unipi.it}
\affiliation{
  \institution{University of Pisa}
  \country{}
}

\begin{abstract}
Neural information retrieval systems typically use a cascading pipeline, in which a first-stage model retrieves a candidate set of documents and one or more subsequent stages re-rank this set using contextualized language models such as BERT. In this paper, we propose \multibert{}, a new document term-weighting scheme suitable for efficient retrieval using a standard inverted index. Compared to existing methods, \multibert{} improves impact-score modeling and tackles the vocabulary-mismatch problem. In particular, \multibert{} leverages \doctfquery{} to enrich the document collection and, using a contextualized language model, directly estimates the semantic importance of tokens in a document, producing a single-value representation for each token in each document. Our experiments show that \multibert{} significantly outperforms prior first-stage retrieval approaches by up to $17\%$ on effectiveness metrics w.r.t. \doctfquery, and, when deployed in a re-ranking scenario, can reach the same effectiveness of state-of-the-art approaches with up to $5.1\times$ speedup in efficiency.
\end{abstract}

\maketitle

\input{sections/1.intro}

\input{sections/2.method}
\input{sections/3.evaluation}

\input{sections/4.conclusion}

\small \textbf{Acknowledgments:} This research was partially supported by NSF Grant IIS-1718680 and CAREER grant CNS-1651570, affiliate members and other supporters of the Stanford DAWN project---Ant Financial, Facebook, Google, Infosys, NEC, and VMware---as well as Cisco and SAP, the Italian Ministry of Education and Research (MIUR) in the framework of the CrossLab project (Departments of Excellence), and by the University of Pisa in the framework of the AUTENS project (Sustainable Energy Autarky). We would like to thank Matei Zaharia for insightful discussions and feedback. Any opinions, findings, and conclusions or recommendations expressed in this material are those of the authors and do not necessarily reflect the views of the National Science Foundation.

\clearpage
\bibliographystyle{acm}
\bibliography{sigir2021}

\end{document}

%% file: sections/1.intro.tex
\section{Introduction}
Modern search engines employ complex, machine-learned ranking functions to retrieve the most relevant documents for a query. Recently, the development of pre-trained contextualized language models such as BERT~\cite{devlin2018bert} has resulted in impressive benefits in search effectiveness, at the cost of expensive query processing times, which can make their deployment in production scenarios challenging. \citet{nogueira2019bertranker} and \citet{cedr} showed the superior performance of BERT in term of effectiveness for passage and document re-ranking tasks, respectively, by fine-tuning the pre-trained transformer network to distinguish between relevant and non-relevant query--document pairs.
However, several recent studies~\cite{cedr,Hofsttter2019LetsMR} have shown that this can have very high computational cost, even if re-ranking just the top 1000 results.
Other studies~\cite{epic,prettr,colbert} proposed methods with lower lower computational cost but typically some loss in retrieval quality. BERT's Transformer encoder is composed of many neural layers performing expensive processing to compute the query-document relevance signals. Different solutions have been proposed to address this performance bottleneck, based on the pre-computation of query-document representations produced by BERT. \epic~\cite{epic} proposes to build on top of BERT a new ranking model trained to generate query and document representations in a given fixed-length vector space, equal to the size of the lexicon. Document representations are pre-computed, while query representations are computed at retrieval time, and then used to obtain a ranking score by computing a similarity between the two representations. %

\prettr~\cite{prettr} and \colbert~\cite{colbert} experimentally show that the query-document interactions in most layers of BERT have little impact on the final effectiveness. This leads them to pre-compute document representations at indexing time, which are used at query processing time to compute the query-document interaction only in a final layer. While \prettr\ still relies upon a first-stage candidate generation based on BM25, \colbert investigates the ability of the pre-computed document representations to identify relevant documents among \textit{all} documents in the index. Due to space/time requirements of the document representation, \colbert leverages approximate nearest neighbor (ANN) search applied to dense representations as a first-stage retrieval system, followed by an exact re-ranking stage, while similar approaches using exact nearest neighbor search~\cite{xiong2020approximate} can perform processing in a single stage.

Following a different paradigm, \citet{deepct} investigated the use of the contextual word representations from BERT to generate more effective document term weights for bag-of-words retrieval. \deepct~\cite{deepct}, for passages, and  \hdct~\cite{deepct2}, for documents,  estimate a term's context-specific importance in each passage/document, by projecting each word’s BERT representation into a single term weight. These term weights are then transformed into term frequency-like integer values that can be stored in an inverted index to be used with classical retrieval models. A main limitation of \deepct\ that we address in this work is that it is trained as a \textit{per-token regression task}, in which a ground truth term weight for every word is needed, and which does not permit the individual impact scores to co-adapt for the downstream objective of identifying relevant documents. %

By storing new integer values as term frequencies in the inverted index, \deepct and \hdct enrich a document's bag-of-words representation with additional document-level context information, to match queries more accurately. Using a different approach, \citet{docTTTTTquery} propose \doctfquery{}, a document expansion strategy to enrich each document with additional terms able to improve the retrieval effectiveness of documents w.r.t. queries for which they are relevant. \doctfquery{} trains a sequence-to-sequence model to predict queries potentially relevant to a given document, and appends these queries to the documents before indexing. As another way of expanding documents, the very recent {\sf SparTerm}~\cite{bai2020sparterm} method predicts an importance score for every term in the vocabulary and uses a gating mechanism to only keep a sparse subset of those, using them to learn an \textit{end-to-end} score for relevant and non-relevant documents. However, this only increases the MRR@10 of \doctfquery{} from 0.277 to 0.279.

\looseness -1 We propose \multibert, a more effective approach for learning a relevance score contribution for term-document pairs that can also be stored in a classical inverted index. \multibert{} improves impact-score modeling and tackles the vocabulary-mismatch problem~\cite{zhao2012modeling} between queries and documents. Instead of learning \textit{independent} term-level scores without taking into account the term co-occurrences in the document, as in \deepct, or relying on unchanged BM25 scoring, as in \doctfquery, \multibert\ directly optimizes the \textit{sum} of query term impacts to maximize the score difference between relevant and non-relevant passages for the query.
In other words, while \deepct learns the term frequency component of existing IR models, e.g., BM25, \textit{in this work we aim at learning the final term impact jointly across all query terms occurring in a passage}.
In this way, our proposed model learns richer interaction patterns among the impacts, when compared to training each impact in isolation. To address vocabulary mismatch, \multibert\ leverages \doctfquery\ to enrich every document  with new terms likely to occur in queries for which the document is relevant. Using a contextualized language model, it directly estimates the semantic importance of tokens in a document, producing a single-value representation for each token in each document that can be stored in an inverted index for efficient retrieval.
Our experiments show that \multibert\ significantly outperforms prior first-stage retrieval approaches by up to $17\%$ on effectiveness metrics w.r.t. \doctfquery. When deployed in a re-ranking scenario, it reaches the same effectiveness as state-of-the-art approaches up to $5.1\times$ faster.

In summary, this paper makes the following contributions:
\vspace{-0.4em}
\begin{itemize}
    \item We propose \multibert{}, a more effective scheme for \textit{jointly learning} term impacts over \textit{expanded documents}.
    \item We evaluate \multibert{} on the MS MARCO passage ranking task.
    We find that \multibert{} can improve ranking effectiveness for passage ranking versus prior first-stage retrieval approaches and is competitive when compared to complex systems based on ANN search, while exhibiting much lower computational costs.
    \item We evaluate \multibert\ as a first-stage model in a re-ranking pipeline, and show that this pipeline matches or outperforms strong baseline approaches, while being highly efficient.
\end{itemize}

%% file: sections/2.method.tex
\section{Deep Impact Framework}

\boldparagraph{Document Expansion.}
In our approach, we leverage \doctfquery\ document expansion to enrich the original document collection with expansion terms. As noted by \citet{doc2query}, document expansion can be seen as a two-fold approach. By adding terms that are already part of the document, it rewrites their frequencies, similar to \deepct. Furthermore, it injects into the passage new terms, originally not part of the document, in order to address the term mismatch problem. We refer to the two as \textit{Rewrite} and \textit{Inject}, respectively. 
Table~\ref{tab:dtfq} summarizes 
the effect of \doctfquery\ when applied to the MSMARCO passage ranking collection, and isolates the two contributions.
While \textit{Rewrite} alone
achieves stronger MRR@10 than \textit{Inject}, the latter achieves higher recall. Using both significantly outperforms either one on both measures. Indeed, \textit{Inject} is important for capturing additional results, but \textit{Rewrite} is needed to then properly weight the injected terms. However, the comparison of \textit{Rewrite} vs. \deepct indicates that \doctfquery\ is still sub-optimal in determining the right frequencies, and resulting impact scores, for the terms.

This motivates our approach, \multibert, where we first use the \textit{Inject} step of \doctfquery\ to add new terms, and then directly learn the right impact scores for both old and newly injected terms.

\begin{table}[h]
\renewcommand{\arraystretch}{0.7}
\caption{Different contributions to effectiveness metrics of \doctfquery\ on the MSMARCO passage ranking collection.}
\centering
\begin{tabular}{l ccccc}
\toprule
& \multirow{2}{*}{\bmtf} & \multirow{2}{*}{\deepct} & \multicolumn{3}{c}{\doctfquery} \\
\cmidrule(lr){4-6}
& &  &\multicolumn{1}{c}{Cumulative} & \multicolumn{1}{c}{\textit{Rewrite}} & \multicolumn{1}{c}{\textit{Inject}}   \\
\midrule
MRR@10 & 0.188 & 0.244 & 0.278 & 0.215 & 0.194 \\ 
Recall & 0.858 & 0.910 & 0.947 & 0.878 & 0.912 \\ 
\bottomrule
\end{tabular}
\label{tab:dtfq}
\end{table}

\begin{figure}
\includegraphics[width=0.7\linewidth]{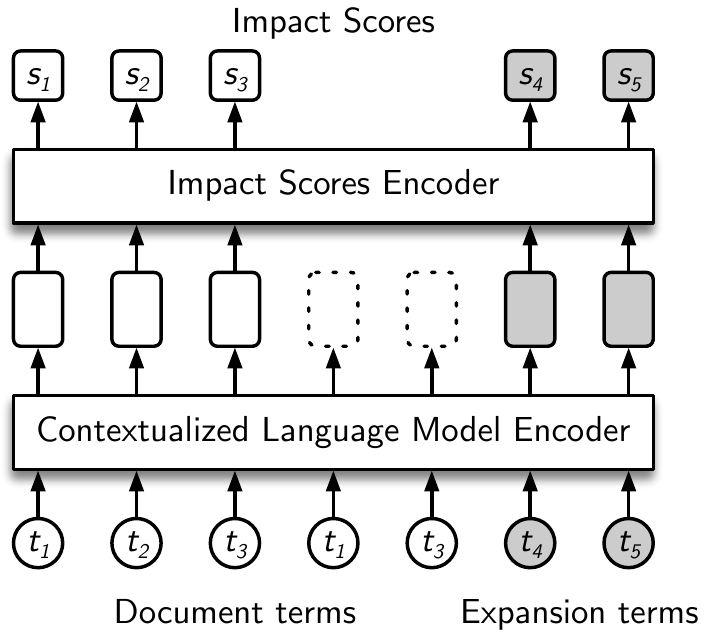}
\caption{Neural network architecture of \multibert.}\label{fig:architecture}

\vspace{-3mm}
\end{figure}

\boldparagraph{Neural Network Architecture.}
\looseness -1 The overall architecture of the \multibert\ neural network is depicted in Figure~\ref{fig:architecture}. \multibert{} feeds a contextual LM encoder the original document terms (in white) and the injected expansion terms (in gray), separating both by a \texttt{[SEP]} separator token to distinguish both contexts. The LM encoder produces an embedding for each input term. The first occurrence of each \textit{unique} term is provided as input to the impact score encoder, which is a two-layer MLP with ReLU activations. This produces a single-value score for each unique term in the document, representing its impact. Given a query $q$, we model the score of document $d$ as simply the sum of impacts for the intersection of terms in $q$ and $d$.

\boldparagraph{Network Training.}
We train our model using triples sampled from the official MS-MARCO training dataset, consisting of a query, a relevant passage, and a presumed non-relevant passage per sample. We expand each passage using the \doctfquery\ as discussed. 
The model converts each document into a list of scores, corresponding to the document terms matching the query. These scores are then summed up, obtaining an accumulated query-document score. For each triple, two scores for the corresponding two documents are computed. The model is optimized via pairwise softmax cross-entropy loss over the computed scores of the documents.  We use BERT-base as the contextualized language model. Max input text length was set to 160 tokens. Losses are back-propagated through the whole \multibert\ neural model with a learning rate of $3 \times 10^{-6}$ with the Adam optimizer. We used batches of 32 triples and train for 100,000 iterations.

\boldparagraph{Impact Scores Computation.}
Following the training phase, \multibert\ can leverage the learned term-weighting scheme to predict the semantic importance of each token of the documents without the need
for queries. Each document is represented as a list of term-score pairs, which are converted into an inverted index. The index can then be deployed and searched as usual for efficient query processing. We infer the scores using three digits of precision, and we do not perform any scaling.

\boldparagraph{Quantization and Query Processing.}
In our approach we predict real-valued document-term scores, also called impact scores, that we store in the inverted index. Since storing a floating point value per posting would blow up the space requirements of the inverted index, we decided to store impacts in a quantized form. The quantized impact scores belong to the range of $[1,2^b-1]$, where $b$ is the number of bits used to store each value. We experimented with $b = 8$ using linear quantization, and did not notice any loss in precision w.r.t. the original scores. Since we quantized all the scores in the index in the same way, to compute a query-document score at query processing we can just sum up all the quantized scores of the document terms matching the query. %

%% file: sections/3.evaluation.tex
\section{Experimental Results}
In this section, we analyze the performance of the proposed method with an extensive experimental evaluation in a realistic and reproducible setting, using state-of-the-art baselines and a standard test collection and query logs.

\boldparagraph{Hardware.}
To evaluate the latency, we use a single core of a machine
with four Intel Xeon Platinum 8268 CPUs and 369 GB of RAM, running Linux 4.18.  
To run \colbert\ a GPU is required, and we used an NVIDIA RTX8000 with 48GB of memory.

\boldparagraph{Dataset and query logs.}
We conduct our experiments on the MSMARCO passage ranking \cite{msmarco} dataset.
To evaluate query processing effectiveness and efficiency, we compare with existing methods using the \msmarcodev{},\footnote{We have made a submission to the official leaderboard and obtained an MRR@10 of 0.318 on the ``eval'' queries.} and we test all methods on the \trecdl{}~\cite{trec2019} and \trecdltw{}~\cite{trec2020} queries from the TREC Deep Learning passage ranking track.

\boldparagraph{Baselines.}
We perform two different sets of experiments. Our initial experiment aims at comparing the performance of \multibert\ as a first-stage ranker, processing queries on inverted indexes but without complex reranking. In this experiment we compare our proposed \multibert\ with the classical \bmtf\ relevance model over the unmodified collection, and state-of-the-art solutions dealing with inverted indexes, namely \deepct, and \bmtf\ over a collection expanded with \doctfquery.
We do not compare with \deepct\ over the collection expanded with \doctfquery, since that would involve training a new \deepct{} model from scratch to learn how to weigh expanded documents.
Our second set of experiments compares \multibert\ in a re-ranking setting. First, the top 1000 documents retrieved by \multibert\ are re-ranked by \epic\ and \colbert and compared to \colbert end-to-end (E2E) where the candidates are generated using ANN search. Finally, we look at first-stage recall and re-ranking-stage MRR@10 when applying \colbert\ at several first-stage cutoffs to different candidate generation methods.

\boldparagraph{Implementations.}
We use Anserini~\cite{anserini} to generate the inverted indexes of the collections. We then export the Anserini indexes using the CIFF common index file format \cite{ciff}, and process them with PISA~\cite{pisa} using the MaxScore query processing algorithm~\cite{maxscore}. 
We use the \bmtf{} scoring method provided by Anserini.
For \deepct{}, we  used the source code and data\footnote{\url{https://github.com/jmmackenzie/term-weighting-efficiency}} provided by \citet{deepct-efficiency}.
For \doctfquery{} we use the predicted queries available online\footnote{https://github.com/castorini/docTTTTTquery}, using 40 concatenated predictions for each passage in the corpus, as recommended by \citet{docTTTTTquery}.
We use the \epic{} implementation in OpenNIR~\cite{onir} and the official pretrained model\footnote{\url{https://github.com/Georgetown-IR-Lab/epic-neural-ir}}.
We use the \colbert{} implementation\footnote{\url{https://github.com/stanford-futuredata/ColBERT}} provided by \citet{colbert}, trained for 200k iterations.
Both training and indexing tasks of \multibert\ are implemented in Python. After the quantization step, the documents are indexed directly by PISA.  
Query processing efficiency is measured using PISA for all baselines. Query processing is performed using MaxScore to retrieve the top $1000$ documents. %
Our source code is publicly available\footnote{\url{https://github.com/DI4IR/SIGIR2021}}. 

\boldparagraph{Metrics.}
To measure effectiveness, we use the oﬃcial metrics for each query set, mean reciprocal rank (MRR@10) for MSMARCO queries, and normalised discounted cumulative gain (NDCG@10) as well as mean average precision (MAP) for TREC queries, following~\cite{hofstatter2020local}. We also report recall  on the first stage and MRR@10 on the re-ranking stage at different cutoff values. Finally, we compute the mean response time (MRT) for every query processing strategy, in ms.
We conduct Bonferroni corrected pairwise t-tests, and report significance with $p < 0.05$.

\boldparagraph{Overall comparison.}
Our first experiment aims to show the early-stage effectiveness improvements that \multibert\ achieves when compared to prior work.
The results are presented in Table~\ref{tab:overall}, which shows effectiveness and efficiency for the three query logs on MS MARCO. We retrieve the top $1000$ documents for each query, without re-ranking, and report the values of NDCG@10, MRR@10, and MAP, as well as MRT. 

\multibert\ significantly outperforms all methods and is statistically significantly better than other strategies for all effectiveness metrics on the \msmarcodev. For the \trecdl\ and \trecdltw\ queries, \multibert\ is always better than the competitors, with statistically significant improvements on NDCG@10 and MAP in some cases. Statistical significance on the latter two query traces is limited by their relatively small number of queries. 

We also see that \multibert\ mean response time exceeds the time reported for other methods. We trace this to the query processing strategy: the distribution of scores induced by BM25, used in \bmtf, \deepct, and \doctfquery\ is exploited more efficiently by the MaxScore algorithm. In contrast, \multibert\ learns new scores, whose distribution is not efficiently exploited by MaxScore. We performed additional experiments using disjunctive query processing without optimizations, omitted for space limitations. These experiments show \multibert\ to be in line with the speed of the other approaches. Optimizing the query processing speed of \multibert{} is an interesting open problem for future research.

\input{figures/effectiveness}

\input{figures/reranking}

\boldparagraph{Re-ranking evaluation.}
Table~\ref{tab:reranking} shows the effect of re-ranking the top $1000$ candidates produced by \multibert\ using two complex re-rankers, \epic\ and \colbert. The table also shows, as a comparison, the performance obtained by \colbert\ when used end-to-end by employing ANN search as the first-stage retrieval mechanism. \multibert\ followed by a \colbert\ re-ranker obtains higher effectiveness values than \colbert\ {\sf E2E} on all query sets. Moreover, \multibert\ + \colbert\ exhibits a $4.4\times - 5.1\times$ speedup w.r.t. \colbert\ {\sf E2E}.

\boldparagraph{First-stage cutoff evaluation.}
\multibert\ is able to achieve statistically significant higher recall than all the compared methods (with one single exception at cutoff 1000). In particular, Table~\ref{table:recall} shows that the gap with the other methods is greater with smaller cutoff values, which reduces the re-ranking cost and thus could enable the use of more complex pairwise ranking models, such as DuoBERT \cite{duobert}. In re-ranking, \multibert\ outperforms all other methods at cutoff 10. Moreover, it outperforms \deepct on all cutoff values except 1000, and it is comparable with \doctfquery.

\input{figures/recall}

%% file: figures/effectiveness.tex
\begin{table}[h]
\renewcommand{\arraystretch}{0.7}
\caption{Effectiveness metrics and mean response time (MRT, in ms) for first-stage methods, on \msmarcodev, \trecdl queries, and \trecdltw queries. The symbol $\triangledown$ denotes a significant difference viz. \multibert}
\centering
\begin{tabular}{lcccc}
\toprule
Strategy & \multicolumn{1}{c}{NDCG@10} & \multicolumn{1}{c}{MRR@10} & \multicolumn{1}{c}{MAP} & \multicolumn{1}{c}{MRT} \\ 
\midrule
\multicolumn{5}{c}{\msmarcodev} \\
\midrule
\bmtf       & 0.235$^\triangledown$ & 0.188$^\triangledown$ & 0.196$^\triangledown$ & 13.24 \\ 
\deepct     & 0.298$^\triangledown$ & 0.244$^\triangledown$ & 0.252$^\triangledown$ & 10.91  \\ 
\doctfquery & 0.338$^\triangledown$ & 0.278$^\triangledown$ & 0.286$^\triangledown$ & 12.62  \\ 
\multibert  & \textbf{0.385}        & \textbf{0.326}        & \textbf{0.332}        & 58.64 \\
\midrule
\multicolumn{5}{c}{\trecdl}  \\
\midrule
\bmtf       & 0.497$^\triangledown$ & 0.683 & 0.290$^\triangledown$ & 10.27 \\ 
\deepct     & 0.578$^\triangledown$ & 0.714 & 0.329$^\triangledown$ & 11.02 \\ 
\doctfquery & 0.648                 & 0.799                 & 0.405 & 11.76 \\ 
\multibert  & \textbf{0.695}        &\textbf{ 0.863}        & \textbf{0.456} & 51.23 \\
\midrule
\multicolumn{5}{c}{\trecdltw}  \\
\midrule
\bmtf       & 0.483$^\triangledown$ & 0.659$^\triangledown$ & 0.286$^\triangledown$ & 14.67  \\ 
\deepct     & 0.550$^\triangledown$ & 0.705 & 0.349$^\triangledown$ & 12.00  \\ 
\doctfquery & 0.619 & 0.742 & 0.408 & 15.51 \\ 
\multibert  & \textbf{0.651} & \textbf{0.820} & \textbf{0.426} & 58.00 \\
\bottomrule
\end{tabular}
\label{tab:overall}

\vspace{-3mm}
\end{table}

%% file: figures/reranking.tex
\begin{table}[h]
\vspace{-3mm}
\renewcommand{\arraystretch}{0.7}

\caption{Effectiveness metrics and mean response time (MRT, in ms) using several re-ranking techniques on \msmarcodev, \trecdl queries, and \trecdltw queries. $\triangledown$ denotes a significant difference viz. \colbert \textsf{E2E}}
\centering
\begin{tabular}{lccc}
\toprule
Strategy & \multicolumn{1}{c}{NDCG@10} & \multicolumn{1}{c}{MRR@10} & \multicolumn{1}{c}{MRT}  \\
\midrule
\multicolumn{4}{c}{\msmarcodev} \\
\midrule
\multibert\ + \epic        & 0.367$^\triangledown$ & 0.303$^\triangledown$ & 194.64 \\
\multibert\ + \colbert     & \textbf{0.425} & \textbf{0.362} & 81.00 \\
\colbert\ \textsf{E2E}     & 0.424 & 0.361 & 380.97 \\
\midrule
\multicolumn{4}{c}{\trecdl}  \\
\midrule
\multibert\ + \epic        & 0.711 & \textbf{0.880} & 191.23 \\
\multibert\ + \colbert     & \textbf{0.722} & 0.826 & 73.29 \\
\colbert\ \textsf{E2E}     & 0.694 & 0.826 & 370.98 \\
\midrule
\multicolumn{4}{c}{\trecdltw}  \\
\midrule
\multibert\ + \epic        & 0.646 & 0.773 & 196.00 \\
\multibert\ + \colbert     & \textbf{0.691} & \textbf{0.781} &  79.84 \\
\colbert\ \textsf{E2E}     & 0.676 & 0.776 & 364.82 \\
\bottomrule
\end{tabular}
\label{tab:reranking}

\vspace{-3mm}
\end{table}

%% file: figures/recall.tex
\renewcommand{\arraystretch}{0.7}

\begin{table}[ht]
\vspace{-2mm}
\renewcommand{\arraystretch}{0.7}
\caption{First-stage recall and re-rank-stage MRR@10 using \colbert\ at several first-stage cutoffs for different candidate generation methods  w.r.t. \msmarcodev. The symbol $\triangledown$ denotes a significant difference viz. \multibert}
\label{table:recall}
\centering
\begin{tabular}{ccccc}
\toprule
$k$    & \bmtf & \deepct & \doctfquery & \multibert \\ 
\midrule
 & \multicolumn{4}{c}{{Recall (first stage)}}\\
\midrule
10    &  0.394$^\triangledown$  &  0.484$^\triangledown$  &  0.542$^\triangledown$  &  \textbf{0.584} \\
20    &  0.483$^\triangledown$  &  0.577$^\triangledown$  &  0.649$^\triangledown$  &  \textbf{0.680} \\
200   &  0.739$^\triangledown$  &  0.816$^\triangledown$  &  0.869$^\triangledown$  &  \textbf{0.882} \\
1000  &  0.858$^\triangledown$  &  0.910$^\triangledown$  &  0.947  &  \textbf{0.948} \\
\midrule
 & \multicolumn{4}{c}{{MRR@10 (re-rank stage)}} \\
\midrule
10     &  0.270$^\triangledown$  &  0.302$^\triangledown$  &  0.341$^\triangledown$  &  \textbf{0.350}   \\
20     &  0.299$^\triangledown$  &  0.322$^\triangledown$  &  0.355  &  \textbf{0.357}   \\
200    &  0.343$^\triangledown$  &  0.353$^\triangledown$  &  \textbf{0.361}  &  \textbf{0.361}   \\
1000   &  0.355$^\triangledown$  &  0.360  &  \textbf{0.362}  &  \textbf{0.362}   \\

\bottomrule

\end{tabular}

\vspace{-3mm}
\end{table}

%% file: sections/4.conclusion.tex
\section{Conclusions and future work}
In this paper, we introduced \multibert, a new first-stage retrieval method that leverages a combination of a traditional inverted indexes and contextualized language models for efficient retrieval. By estimating semantic importance, \multibert\ produces a single-value impact score for each tokens of a document collection. Our results show that \multibert\ outperforms every inverted-index based baseline, in some cases even matching the effectiveness of more complex neural retrieval approaches such as \colbert. Furthermore, when \colbert\ is used to re-rank candidates retrieved by \multibert\, instead of approximate nearest neighbor, we find a dramatic reduction of query processing latency, and a more modest improvement in effectiveness of the whole pipeline.  
Future work will focus on further enhancing the underlying model. First, we would like to experiment with more relaxed matching conditions, instead of exact match, between the query-document terms. Second, we believe that we could improve further term expansion with more sophisticated techniques. Finally, we plan to investigate how changing the distribution of impact scores affects query processing algorithms such as MaxScore, and how we can address this issue.